\documentclass[]{spie}  

\usepackage{amsmath,amsfonts,amssymb}
\usepackage{graphicx}
\usepackage[colorlinks=true, allcolors=blue]{hyperref}

\title{Total variation vs L1 regularization: a comparison of compressive sensing optimization methods for chemical detection}

\author[a]{Elin Farnell}
\author[a]{Henry Kvinge}
\author[b]{Julia R. Dupuis}
\author[a]{Michael Kirby}
\author[a]{Chris Peterson}
\author[b]{Elizabeth C. Schundler}
\affil[a]{Colorado State University, Department of Mathematics, 1874 Campus Delivery, Fort Collins, CO 80523-1874, USA}
\affil[b]{Physical Sciences Inc., 20 New England Business Center, Andover, MA 01810-1077, USA}

\authorinfo{Further author information (E-mail): (Send correspondence to E.F.)\\E.F.: elinfarnell@gmail.com;
H.K.: hkving@gmail.com; M.K.: kirby@math.colostate.edu;\\ C.P.: peterson@math.colostate.edu; J.R.D.: jdupuis@psicorp.com; E.C.S.: eschundler@psicorp.com}

\begin{document} 
\maketitle

\begin{abstract}
One of the fundamental assumptions of compressive sensing (CS) is that a signal can be reconstructed from a small number of samples by solving an optimization problem with the appropriate regularization term. Two standard regularization terms are the L1 norm and the total variation (TV) norm. We present a comparison of CS reconstruction results based on these two approaches in the context of chemical detection, and we demonstrate that optimization based on the L1 norm outperforms optimization based on the TV norm. Our comparison is driven by CS sampling, reconstruction, and chemical detection in two real-world datasets: the Physical Sciences Inc. Fabry-P\'{e}rot interferometer sensor multispectral dataset and the Johns Hopkins Applied Physics Lab FTIR-based longwave infrared sensor hyperspectral dataset. Both datasets contain the release of a chemical simulant such as glacial acetic acid, triethyl phosphate, and sulfur hexafluoride. For chemical detection we use the adaptive coherence estimator (ACE) and bulk coherence, and we propose algorithmic ACE thresholds to define the presence or absence of a chemical of interest in both un-compressed data cubes and reconstructed data cubes. The un-compressed data cubes provide an approximate ground truth. We demonstrate that optimization based on either the L1 norm or TV norm results in successful chemical detection at a compression rate of 90\%, but we show that L1 optimization is preferable. We present quantitative comparisons of chemical detection on reconstructions from the two methods, with an emphasis on the number of pixels with an ACE value above the threshold.
\end{abstract}


\section{INTRODUCTION}
\label{sec:intro}  

Hyperspectral imaging is a key tool for problems that require high-precision measurements across various spectral ranges. One such problem is the detection of specific chemicals in a scene of interest. More generally, hyperspectral imaging has been employed for a wide range of applications from food quality assessment to surface composition and mineralogy mapping~\cite{gowen2007hyperspectral,Lorente2012,van2012multi}.

One of the challenges associated with hyperspectral imaging is the high expense associated with sensing outside of the visible range. As a consequence, compressive sensing (CS) has become an important tool for hyperspectral imaging~\cite{willett2014sparsity}. Using the CS framework, it is possible to build a hyperspectral imaging device with just a single sensor (a so-called single-pixel camera)~\cite{baraniuk2007compressive}. Such devices make applications of hyperspectral imagery accessible at a significantly lower cost than traditional models.

In the context of CS, one begins with an ill-posed problem - there are infinitely many scenes that could produce a particular set of CS observations. The traditional approach is to solve a related optimization problem with a regularization term. There are many options for the particular optimization problem and regularization term, with $\ell_1$-regularization and total variation (TV) regularization being standard choices~\cite{baraniuk2007compressive,candes2008introduction,rudin1992nonlinear}. In this paper, we seek to address the question of which of these two approaches is preferable in terms of chemical detection in reconstructed data. We focus specifically on the context of chemical detection in datasets that contain chemical releases or chemical simulant releases. In order to provide quantitative evidence, we simulate CS sampling on data so that we have an approximate ground truth for comparison.  

To measure the accuracy of reconstructed hyperspectral data, we focus on the ability to detect the presence or absence of a target chemical signature in raw and reconstructed data. We use the adaptive coherence estimator (ACE) as our detector algorithm. While we wish to evaluate based on a binary choice between ``chemical present'' and ``chemical absent'' in each pixel, the output of the ACE algorithm is instead a real number between $0$ and $1$. To make this conversion, a threshold is necessary. In order to make a fair comparison between raw and reconstructed data we introduce a method of determining this threshold. To our knowledge this method is new. 

This paper is organized as follows. In Sec.~\ref{sec:background}, we provide background on CS, TV and $\ell_1$-regularization, and methods for chemical detection. In Sec.~\ref{sec:Comparison}, we propose a threshold for use with chemical detection methods that makes objective comparison possible. We then provide results that compare TV and $\ell_1$-regularization as applied to two hyperspectral datasets. We further address robustness of the two regularization approaches to variation in the threshold choice. Finally, we summarize our findings and propose questions for future work in Sec.~\ref{sec:Conclusion}.


\section{BACKGROUND}
\label{sec:background}

In this paper we will generally use upper case letters (for example $A$, $S$, $X$, $H$, and $Y$) to denote matrices. We will use lower case letters (for example $u$, $x$, $\tilde{x}$, and $y$) to denote vectors. Matrices being placed side by side always denotes standard matrix multiplication.

\subsection{Compressive Sensing}
\label{sec:CS}

Compressive sensing (CS) is a collection of methods that allow accurate reconstruction of certain classes of signals even when measurements are low-dimensional linear functionals of raw data~\cite{baraniuk2007compressive}. In the language of linear algebra, CS can be equivalently formulated as being a set of methods for solving the ill-posed problem of finding $\tilde{x}$ from 
\begin{equation} \label{eqn-basic-problem}
y = A\tilde{x} \in \mathbb{R}^k
\end{equation}
when $A$ is a $k \times n$ matrix, $\tilde{x} \in \mathbb{R}^n$, and $k < n$. While in general there will be an infinite number of solutions $x'$ such that $y = Ax'$, additional assumptions about $\tilde{x}$ allow us to choose a close approximation to $\tilde{x}$ among all $x'$. Thus, rather than solving the algebraic equation \eqref{eqn-basic-problem} we solve an optimization problem in which we minimize some functional on all $x'$ satisfying $Ax' = y$. In this way our assumptions about $\tilde{x}$ inform a choice of regularization term. That is, instead of solving \eqref{eqn-basic-problem} we solve some variation of the optimization problem
\begin{equation} \label{eqn-general-optimization-prob}
\underset{{x \in \mathbb{R}^n}} {\text{argmin}}\; \Phi(x) \quad\quad \text{such that} \quad Ax = y,
\end{equation}
where $\Phi(x)$ is a functional measuring the extent to which $x$ satisfies some assumption about $\tilde{x}$.


In our application of CS to hyperspectral imagery
we assume each data cube $X$ has $b$ bands and each band is of size $n$ (that is, as a 2-dimensional array, each band has size $n_1 \times n_2$ where $n_1n_2 = n$). When manipulating hyperspectral data for the experiments in this paper we generally flatten each band so that we can treat the whole hyperspectral cube as a matrix. Hence in our setting $x \in \mathbb{R}^n$ becomes $X \in \mathbb{R}^{n \times b}$ and the problem is to try to reconstruct the $n \times b$ matrix $X$ when we are only given a $k \times b$ matrix $Y$ such that $Y = SX$. Analogously to above, this problem will be solved by making assumptions about the sparsity of the columns of $X$ under certain transformations.

\subsection{Total Variation and L1 Regularization}
\label{sec:TVL1}

Two of the most popular choices for $\Phi(x)$ in \eqref{eqn-general-optimization-prob} are:
\begin{itemize}
    \item {\emph{$\ell_1$-regularization}}: Here $\Phi(x)$ is the $\ell_1$-norm, $\Phi(x) = ||x||_{\ell_1}$. While it is uncommon for images (hyperspectral or otherwise) to be sparse in their natural basis, it is well known that they are very often sparse in other bases such as wavelet bases. If $H,H^{-1}: \mathbb{R}^n \rightarrow \mathbb{R}^n$ are the $1$-dimensional Haar wavelet transform and its inverse, respectively, then to require a solution $x = H^{-1}u$ to have sparsity in the wavelet basis we solve the modified optimization problem:
    \begin{equation} \label{eqn-opt-L1}
        \underset{u \in \mathbb{R}^n}{\text{argmin}}\;||u||_{\ell_1} \quad\quad \text{such that} \quad\quad SH^{-1}u = y,
    \end{equation}
    where $S\in\mathbb{R}^{k\times n}$ is a sampling matrix. Note that informally, this problem is asking us to find $x$ such that when we sample $x$ with $S$ we get $y$, and that $x$ is maximally sparse in the Haar wavelet basis (i.e. $u = Hx$ is maximally sparse). In this paper all experiments are done with respect to the $1$-dimensional Haar wavelet basis. 
    
    The specific version of \eqref{eqn-opt-L1} used to reconstruct a hyperspectral image $X$ from $\tilde{X}$ (given as an $n \times b$ matrix) is 
    \begin{equation}\label{eqn-opt-L1-hyperspectral}
        \underset{U \in \mathbb{R}^{n\times b}}{\text{argmin}}\;||U||_{\ell_1} \quad\quad \text{such that} \quad\quad SH^{-1}U = Y. 
    \end{equation}
    Note that in this case $X = H^{-1}U$, so that we are applying the inverse of a $1$-dimensional Haar wavelet transform to each band independently, taken as a vector (or column) of $U$. Now $Y$ takes the form of a $(k \times b)$ matrix (rather than just a vector) where each column is the set of samples obtained from a specific band (column) of $X$. We define $||U||_{\ell_1}$ to be the usual $\ell_1$-norm applied to $U$ as a length $nb$-vector. 

    \item {\emph{TV-regularization}}: In this case $\Phi(x)$ is the TV functional $\Phi(x) = ||x||_{TV}$. There are two different version of the TV functional. The isotropic TV functional was the version originally proposed \cite{rudin1992nonlinear}. The anisotropic version is slightly easier to minimize and hence we have used this version in this paper. If $x = [x_{ij}]$ is an $n_1 \times n_2$ array, then
    \begin{equation} \label{eqn-tv-def}
        ||x||_{TV} := \sum_{i = 1}^{n_1}\sum_{j = 1}^{n_2} |x_{i+1,j} - x_{i,j}| + |x_{i,j+1} - x_{i,j}|
    \end{equation}
    where indices are taken modulo $n_1$ and modulo $n_2$ respectively. Note that if $\nabla_x$ and $\nabla_y$ are the gradient operators then we can write \eqref{eqn-tv-def} as 
    \begin{equation*}
        ||x||_{TV} = ||\nabla_xx||_{\ell_1} + ||\nabla_y x||_{\ell_1}.
    \end{equation*}
    TV-regularization is also widely used for denoising. One advantage of this global approach is that it is much more likely to preserve features such as edges that would be smoothed by other denoising techniques.
        The optimization problem then solved to reconstruct $x$ in the TV case is:
    \begin{equation} \label{eqn-opt-TV}
        \underset{x \in \mathbb{R}^n}{\text{argmin}}\;||x||_{TV} \quad\quad \text{such that} \quad\quad Sx = y,
    \end{equation}
    where $S$ is again a $k\times n$ sampling matrix. 
   In the case where $X$ is a hyperspectral image, 
    we write
    \begin{equation} \label{eqn-opt-TV-hyperspectral}
        \underset{X \in \mathbb{R}^{n \times b}}{\text{argmin}}\;||X||_{TV} \quad\quad \text{such that} \quad\quad SX = Y.
    \end{equation}
    Here $||X||_{TV}$ is defined as
    \begin{equation} \label{eqn-TV-matrices}
        ||X||_{TV} = ||\nabla_xX||_{\ell_1} + ||\nabla_y X||_{\ell_1}
    \end{equation}
    where the $\ell_1$-norm is applied to the matrices resulting from matrix multiplication in \eqref{eqn-TV-matrices} by flattening them and treating them as length $nb$ vectors.
    
\end{itemize}

Partly driven by interest in CS, efficient methods for solving \eqref{eqn-opt-L1} and \eqref{eqn-opt-TV} have been developed. In this paper we used the split Bregman method \cite{GO09} to solve both \eqref{eqn-opt-L1-hyperspectral} and \eqref{eqn-opt-TV-hyperspectral}.

\subsection{Chemical Detection: ACE and Bulk Coherence}
\label{sec:ACEBC}

In this paper, we present a comparison of TV and $\ell_1$ approaches to CS reconstruction on datasets that contain observations of chemical explosions. Hence, we measure success in terms of chemical detection in reconstructed cubes. In this section, we provide brief descriptions of the standard approaches to chemical detection that we use.

The \emph{adaptive coherence estimator} (ACE)~\cite{scharf1996adaptive,kraut2001adaptive} is a well-known technique used for chemical detection. Let $s$ be a spectral signature for a target chemical and let $x$ be a spectral signature in a specific pixel within a hyperspectral cube (in the literature, $x$ is the \emph{pixel under test} (PUT)). The ACE statistic is the square of the cosine of the angle between $s$ and $x$ relative to the background. To be precise, the ACE statistic is calculated in this setting as 
$$\frac{(s^T\Gamma^{-1}x)^2}{(s^T\Gamma^{-1}s)(x^T\Gamma^{-1}x)},$$
where $\Gamma$ is the maximum likelihood estimator for the covariance matrix of background data.

We additionally use a \emph{bulk coherence} statistic for signal enhancement. This statistic is also called the {\emph{multipulse coherence estimator}} (MPACE)~\cite{pakrooh2017adaptive,scharf2017multipulse}. The motivation for using this statistic is to improve detection when there are neighborhoods of pixels that contain multiple high ACE values. Bulk coherence at a given pixel is defined by incorporating the ACE statistic values in that pixel and neighboring pixels: let $c_i$ be the $i$-th pixel ACE value in a neighborhood of $M$ pixels. Then we compute bulk coherence as 
$$1-\prod_{i=1}^M(1-c_i).$$
If several ACE values in a neighborhood are close to 1, then $\prod_{i=1}^M(1-c_i)$ will be close to zero, thus giving a bulk coherence value near 1. In practice, we use $M=9$ to create a $3\times 3$ neighborhood centered on a pixel to compute bulk coherence.

In our experimental results, we observe improved chemical detection resulting from the use of bulk coherence. We additionally seek to reduce noise via a filter we refer to as \emph{persistence}. Specifically, persistence defines a value for a pixel to be zero if its bulk coherence fails to remain above a threshold for at least five successive time frames.

\section{COMPARISON: TOTAL VARIATION VS L1 REGULARIZATION}
\label{sec:Comparison}

In the context of CS for hyperspectral imagery, an important application is chemical detection. We focus our comparison of TV and $\ell_1$-regularization on this context and we emphasize quantitative aspects of this comparison on real-world datasets that contain chemical simulant releases. We begin by introducing a means of algorithmically-determining thresholds so that we can objectively compare the number of pixels in a hyperspectral cube that are deemed to have a particular chemical present. Then we present results on various datasets to demonstrate comparative performance on cubes reconstructed via the two methods.

\subsection{Thresholds for ACE and MPACE}
\label{sec:AlgThresh}

In this section, we propose a method of setting thresholds for the ACE statistic. By construction, the threshold definition is responsive to the device used for CS, the spectral signature of the chemical of interest, and the reconstruction optimization approach. The methodology described here for ACE thresholds can be applied in the same manner for bulk coherence thresholds. 

We define the ACE threshold by using ACE values for a specific chemical of interest on a set of hyperspectral cubes that have been sampled and which do not contain an observation of that chemical. We call these \emph{background} cubes. The motivation behind our threshold definition is that the threshold should be slightly larger than the ACE values generally observed in background cubes. The precise definition is as follows.

Choose parameters $\alpha$ and $\beta.$ In the results we present in this paper, we use $\alpha=99;$ we use $\beta=2$ for sampled and reconstructed data and $\beta=1$ for raw, uncompressed data. Consider a collection of reconstructed background cubes $B=\{B_1,B_2,\ldots,B_n\}.$ Compute the ACE statistic for a chemical of interest in all pixels in all cubes: let $S_i$ be the set of all ACE values computed in cube $B_i$. For each $B_i,$ define $x_i\in S_i$ to be the smallest element in $S_i$ such that $\alpha$ percent of the pixels in $B_i$ have values below $x_i.$ Then we define the threshold $T$ to be $$T=\beta\cdot \textrm{median}\{x_i\}_{i=1}^n.$$

Based on experimental evidence, this threshold definition appears to effectively capture the presence and absence of chemicals of interest. 

\subsection{Performance: Chemical Detection}
\label{sec:Performance}
We present results from two datasets: the Physical Sciences Inc. Fabry-P\'{e}rot interferometer sensor multispectral dataset and the Johns Hopkins Applied Physics Lab FTIR-based longwave infrared sensor hyperspectral dataset~\cite{cosofret2009airis,broadwater2011primer}. These datasets contain chemical simulant releases and provide an ideal setting for comparison of the two optimization approaches to CS presented in this paper. In both cases, we compare the number of pixels with ACE (bulk coherence) values above a threshold in reconstructed cubes against the number of pixels with ACE (bulk coherence) values above a threshold in uncompressed cubes. By using the algorithmically determined threshold as described in Sec.~\ref{sec:AlgThresh}, we make an objective comparison of the two methods. We present results from two chemical release datasets from the Fabry-P\'{e}rot data and two from the Johns Hopkins data.

All results presented in this paper are from simulated CS on raw data with 90\% compression. We use shifted Walsh-Hadamard sampling with a fixed maximal-variance order~\cite{farnell2019sampling}. In all cases, we restrict to a $64\times 64$ field of view that contains the chemical release.

\begin{figure}[ht]
\begin{center}
\begin{tabular}{c}
\includegraphics[height=5cm]{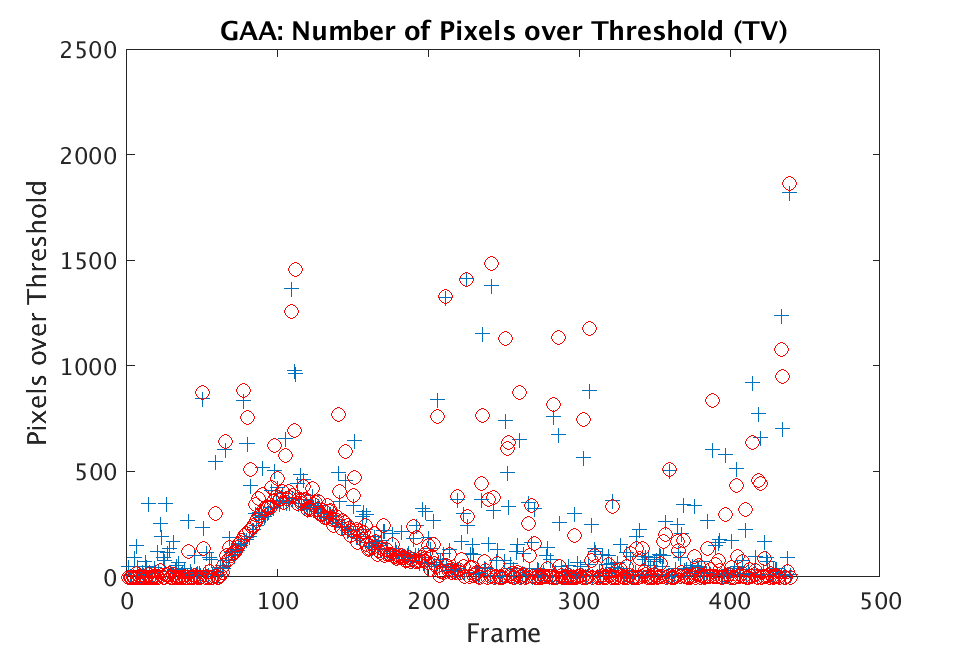}
\includegraphics[height=5cm]{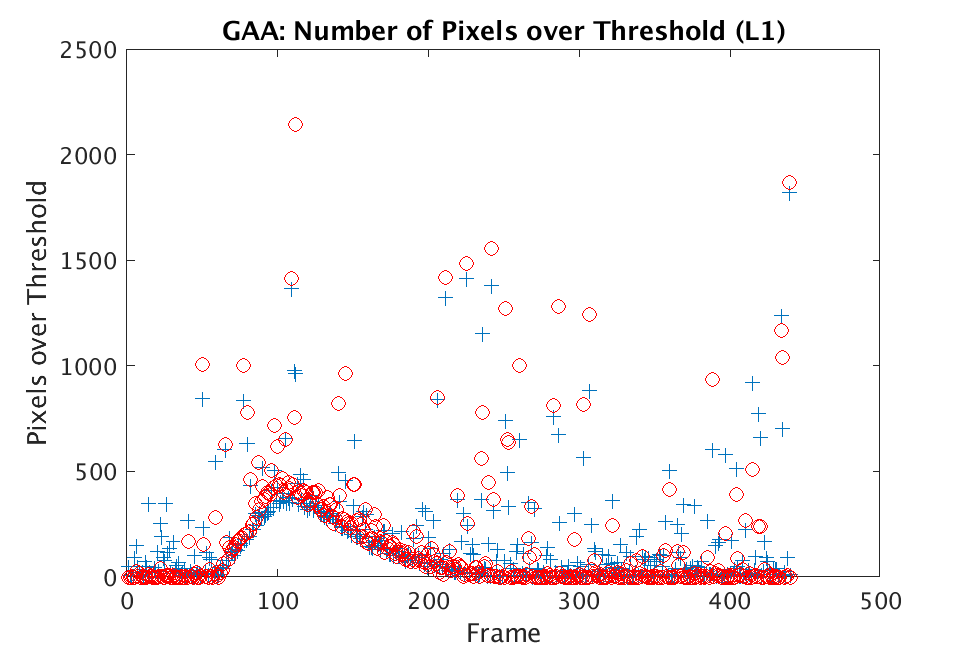}\\
\includegraphics[height=5cm]{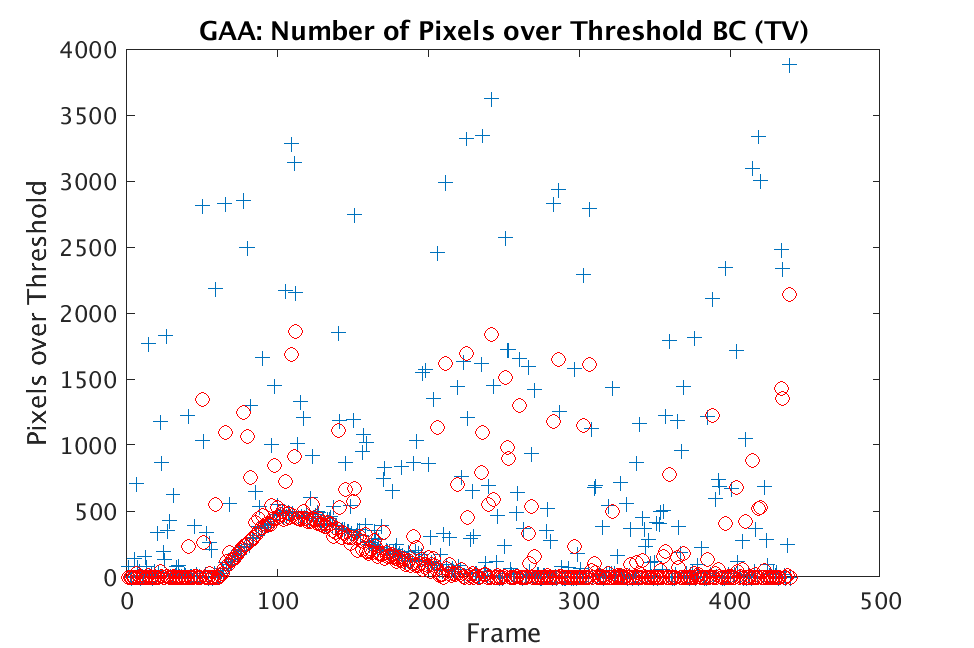}
\includegraphics[height=5cm]{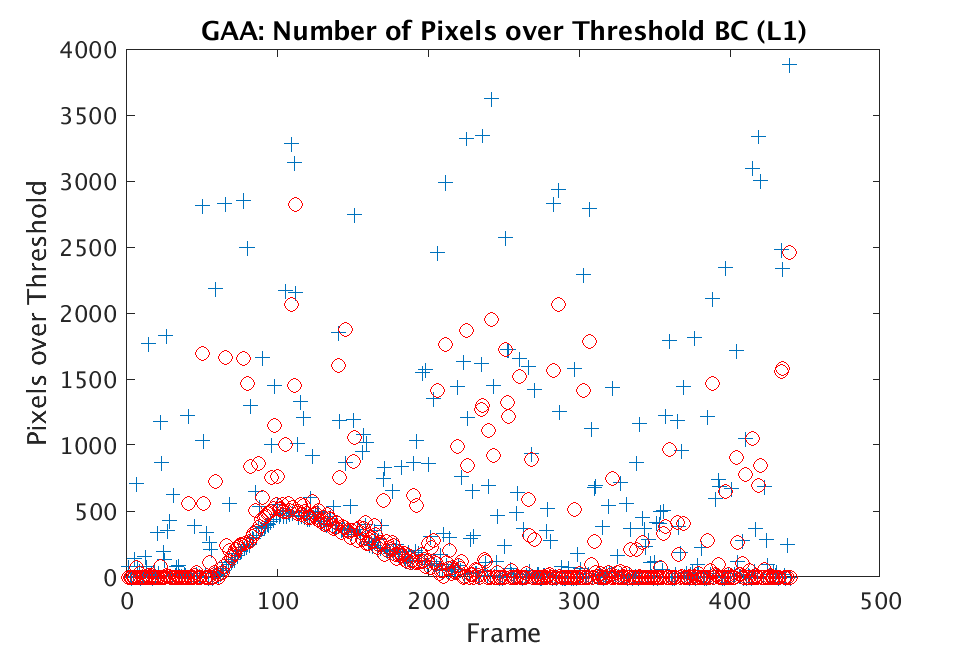}\\
\includegraphics[height=5cm]{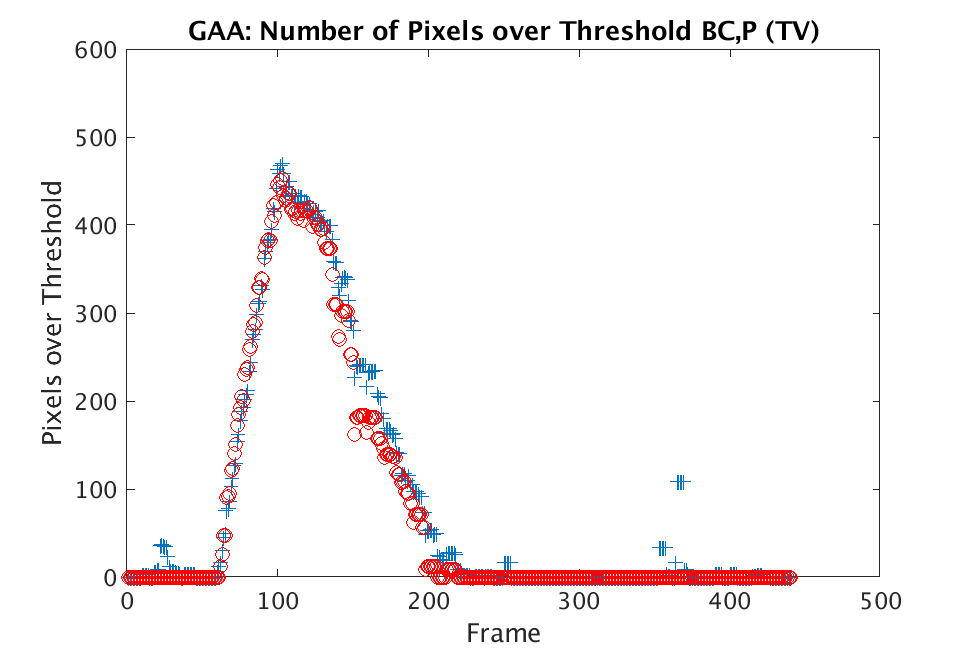}
\includegraphics[height=5cm]{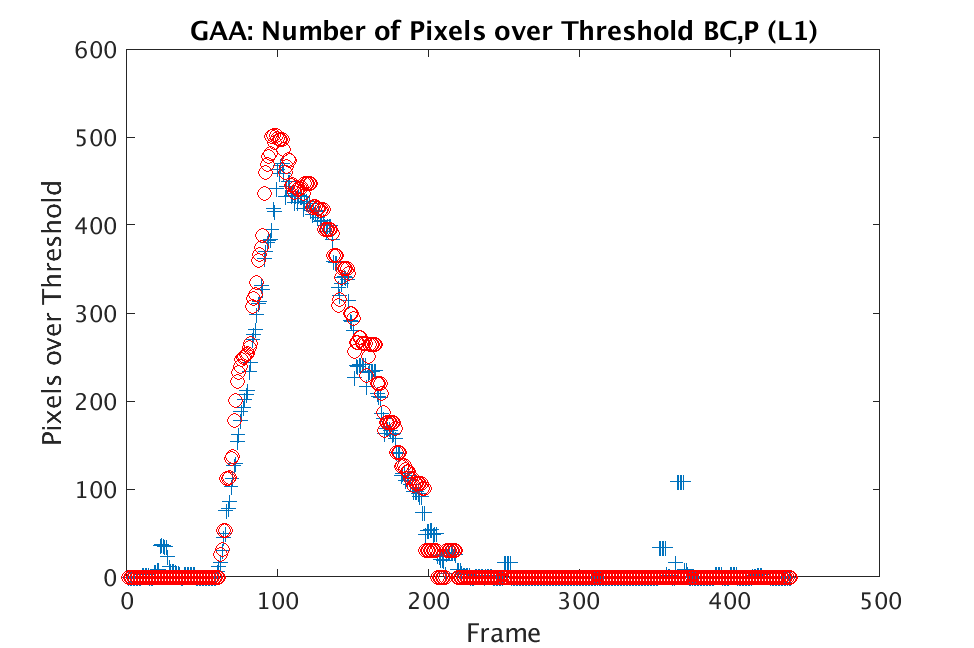}
\end{tabular}
\end{center}
\caption{\label{fig:GAA}} Comparison of the number of pixels with an ACE (bulk coherence) value that exceeds the corresponding threshold (defined as in Sec.~\ref{sec:AlgThresh}) for uncompressed data (blue crosses) and reconstructed data (red circles) as a function of time frame in the Fabry-P\'{e}rot GAA dataset. The figures in the left column are produced from TV reconstructions, and the figures in the right column are from $\ell_1$ reconstructions. The figures are organized by row according to the number of pixels with values over the threshold for (top to bottom): ACE values, bulk coherence values, and bulk coherence with persistence values. All results are computed after sampling at $90\%$ compression. Note that the inclusion of persistence removes a significant amount of noise and makes it clear that both reconstruction approaches result in chemical detection that is consistent with chemical detection on uncompressed data. 
\end{figure}

\begin{figure}[ht]
\begin{center}
\begin{tabular}{c}
\includegraphics[height=5cm]{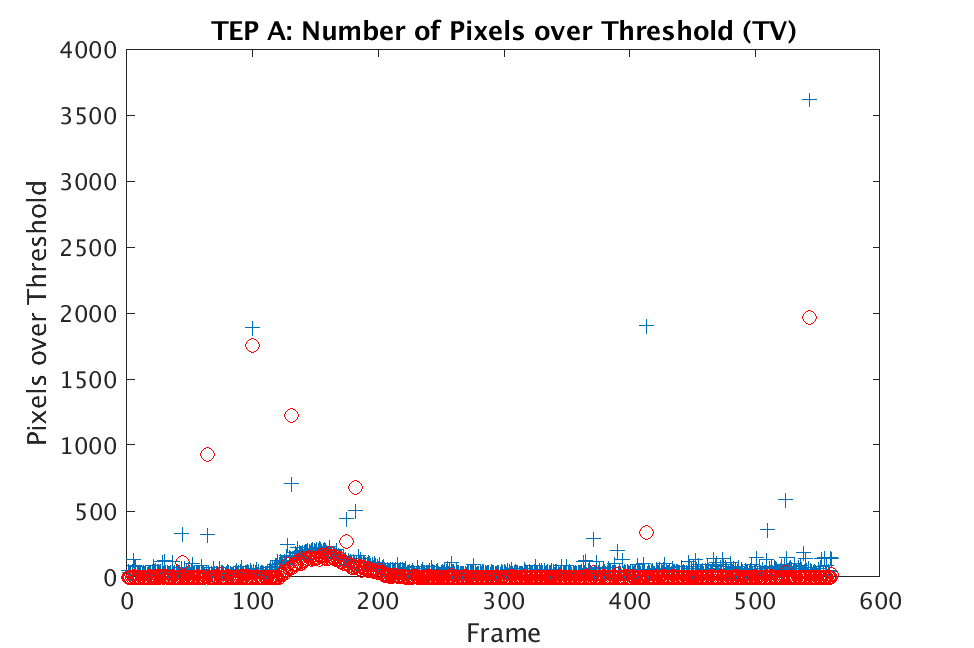}
\includegraphics[height=5cm]{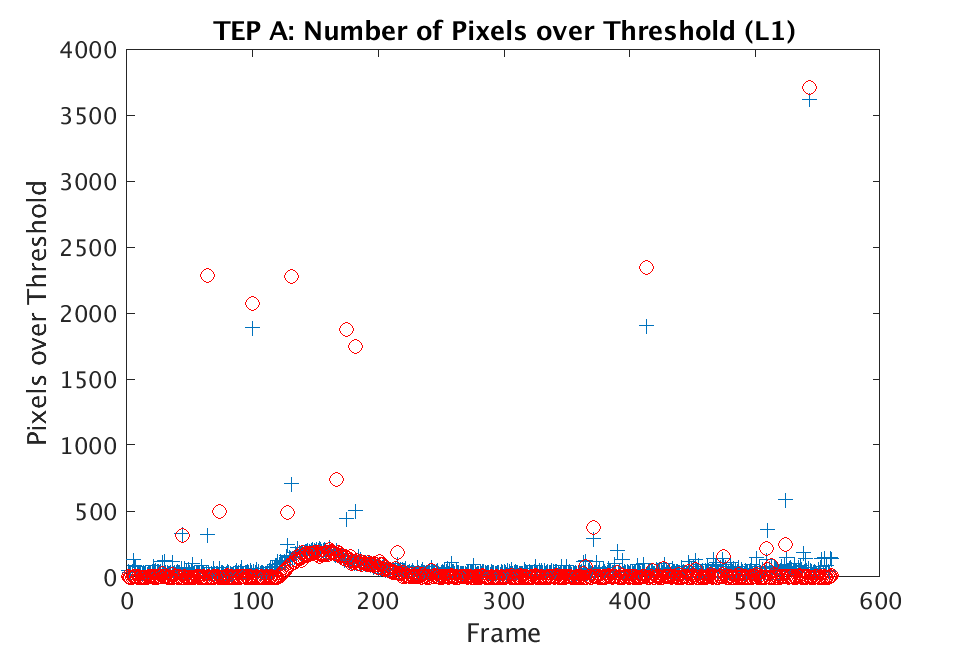}\\
\includegraphics[height=5cm]{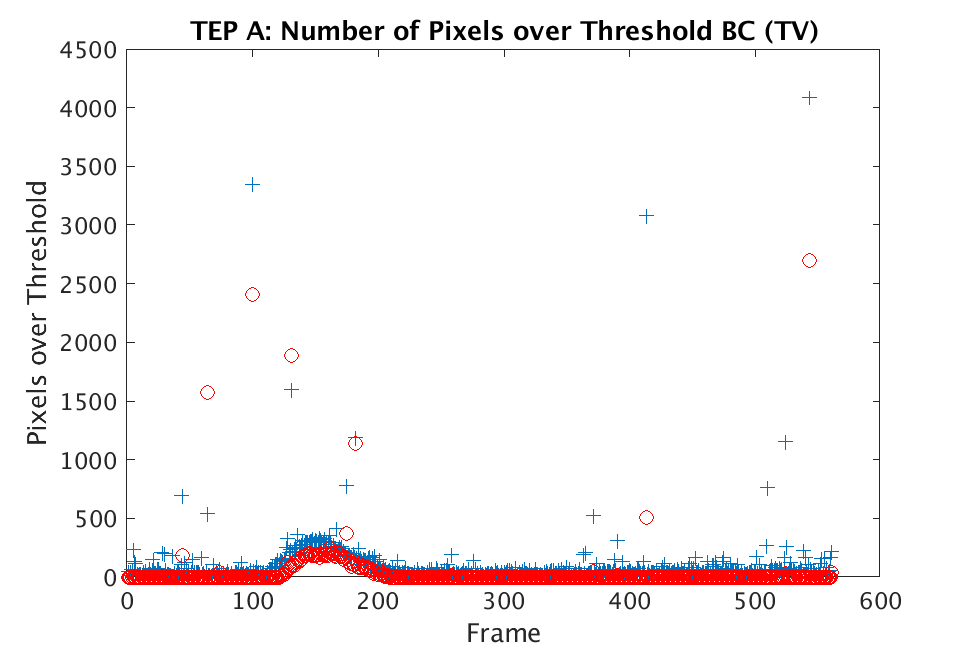}
\includegraphics[height=5cm]{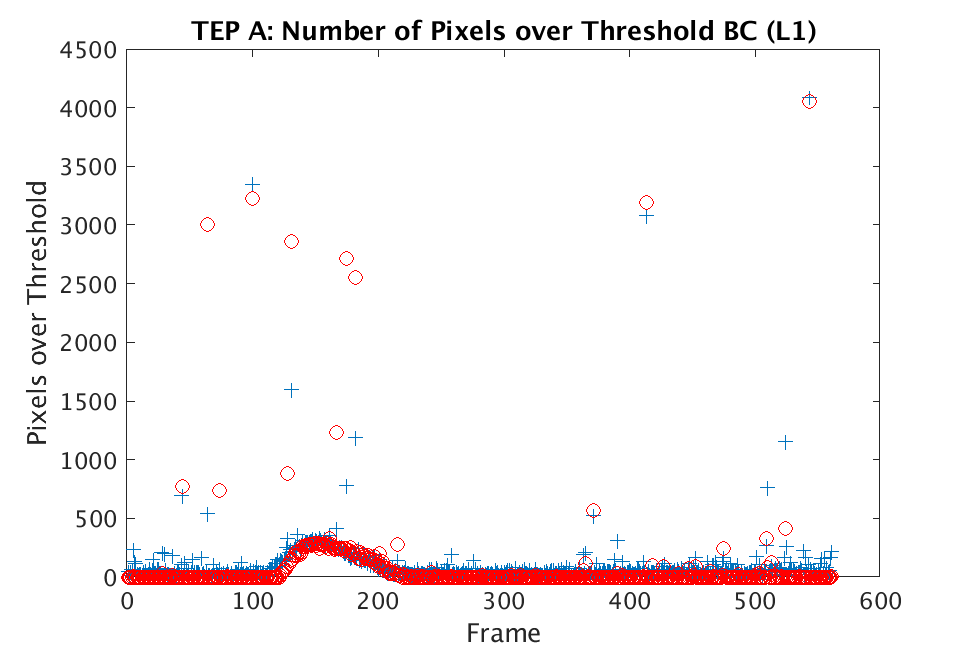}\\
\includegraphics[height=5cm]{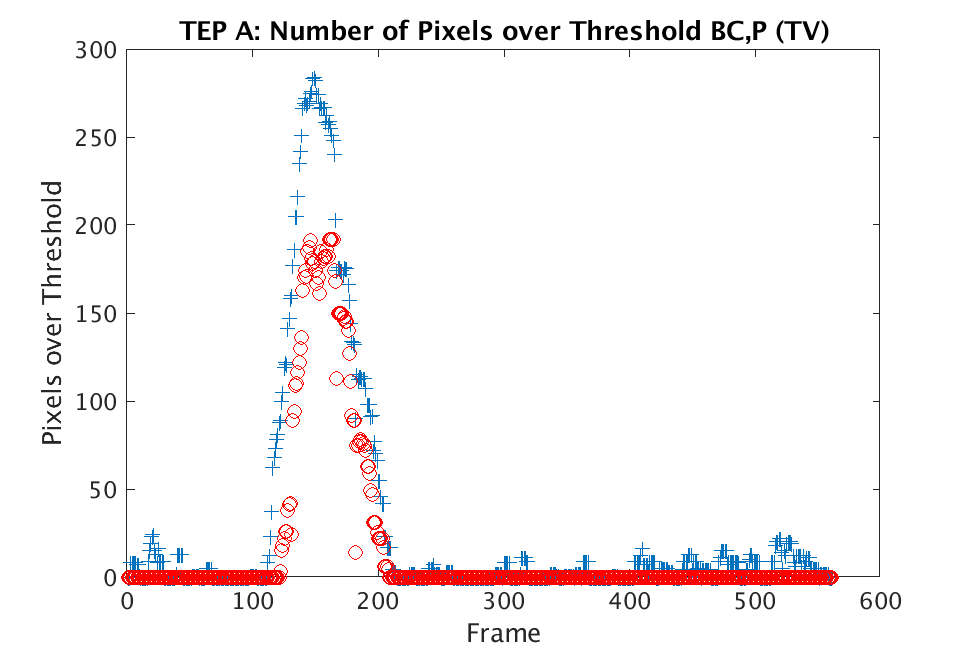}
\includegraphics[height=5cm]{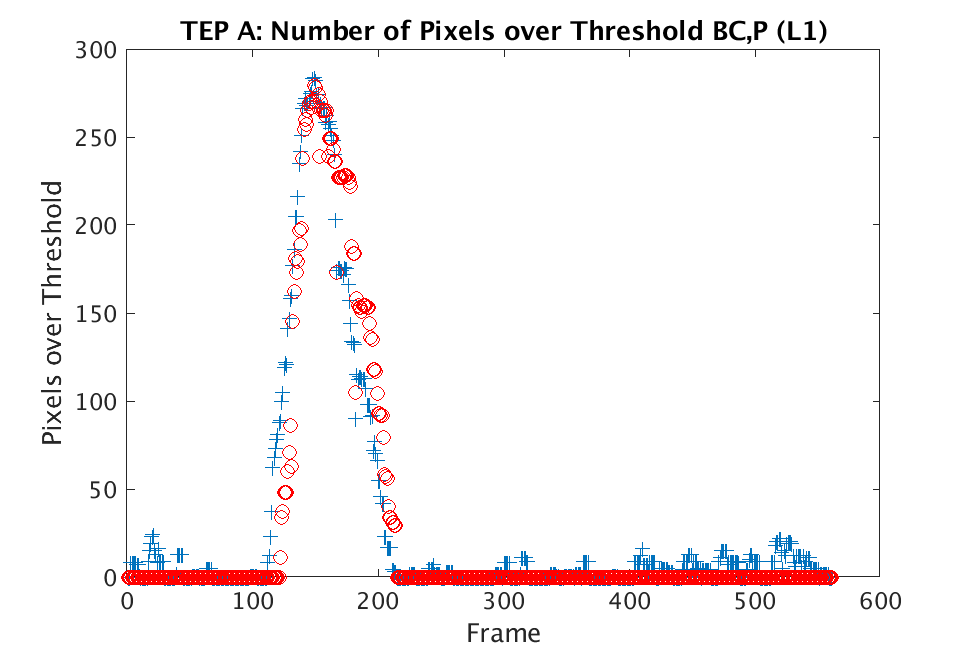}
\end{tabular}
\end{center}
\caption{\label{fig:TEPA}} Comparison of the number of pixels with an ACE (bulk coherence) value that exceeds the corresponding threshold (defined as in Sec.~\ref{sec:AlgThresh}) for uncompressed data (blue crosses) and reconstructed data (red circles) as a function of time frame in the Fabry-P\'{e}rot TEP A dataset. The figures in the left column are produced from TV reconstructions, and the figures in the right column are from $\ell_1$ reconstructions. The figures are organized by row according to the number of pixels with values over the threshold for (top to bottom): ACE values, bulk coherence values, and bulk coherence with persistence values. All results are computed after sampling at $90\%$ compression. In this case, both reconstruction approaches result in effective chemical detection, but the $\ell_1$ reconstruction better reflects the chemical detection as computed on uncompressed data. 
\end{figure}

In Figs.~\ref{fig:GAA} and~\ref{fig:TEPA}, we show the results of chemical detection on Fabry-P\'{e}rot GAA and TEP A datasets. In both figures, blue crosses correspond to the number of pixels over the threshold for uncompressed data, and red circles represent the number of pixels over the threshold for reconstructed data. Total variation results are shown in the left column and $\ell_1$ results are shown in the right column. Finally, the rows represent ACE results, bulk coherence results, and bulk coherence and persistence results, respectively. Notice that in both datasets, there is significant noise that results in spikes for both uncompressed and reconstructed data; the cleanest results appear in the final row where persistence has been implemented. Most importantly, cubes reconstructed with both methods exhibit chemical detection that is consistent with results on uncompressed data. And in both cases, there is slightly stronger detection that occurs on cubes reconstructed with $\ell_1$-regularization.

\begin{figure}[ht]
\begin{center}
\begin{tabular}{c}
\includegraphics[height=5cm]{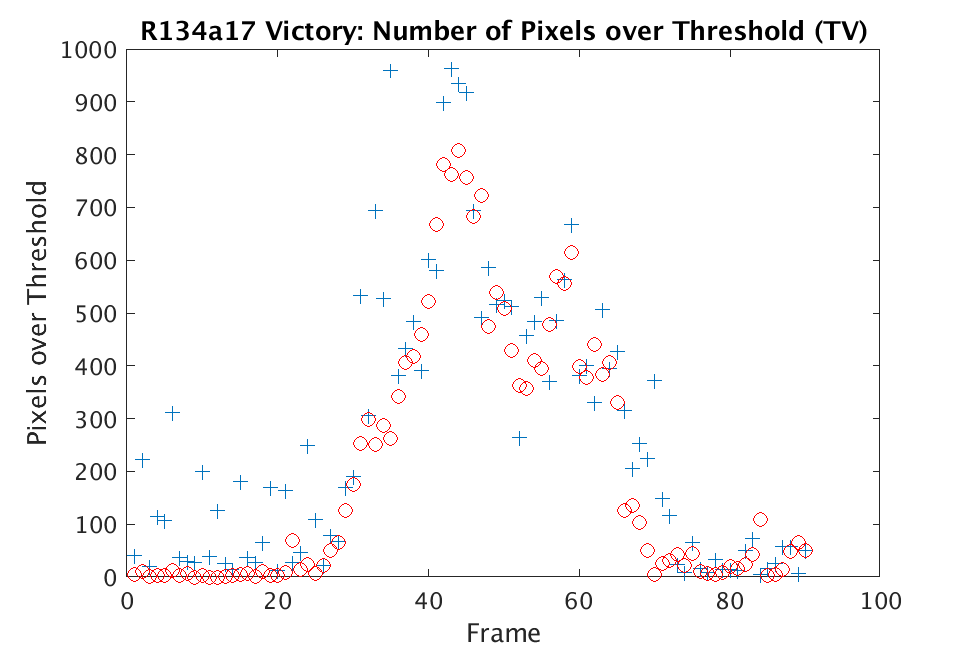}
\includegraphics[height=5cm]{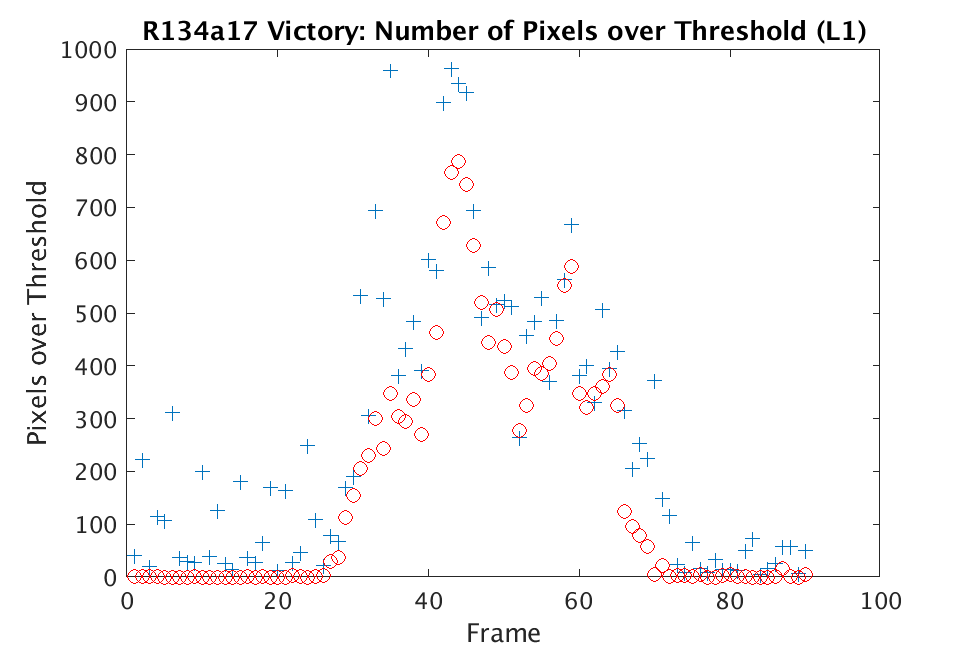}\\
\includegraphics[height=5cm]{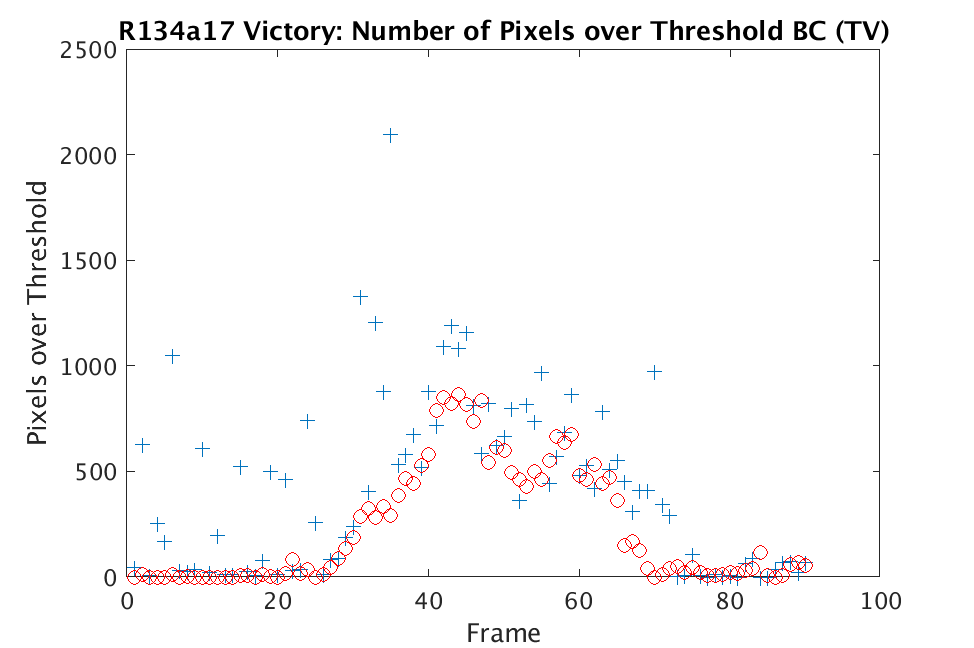}
\includegraphics[height=5cm]{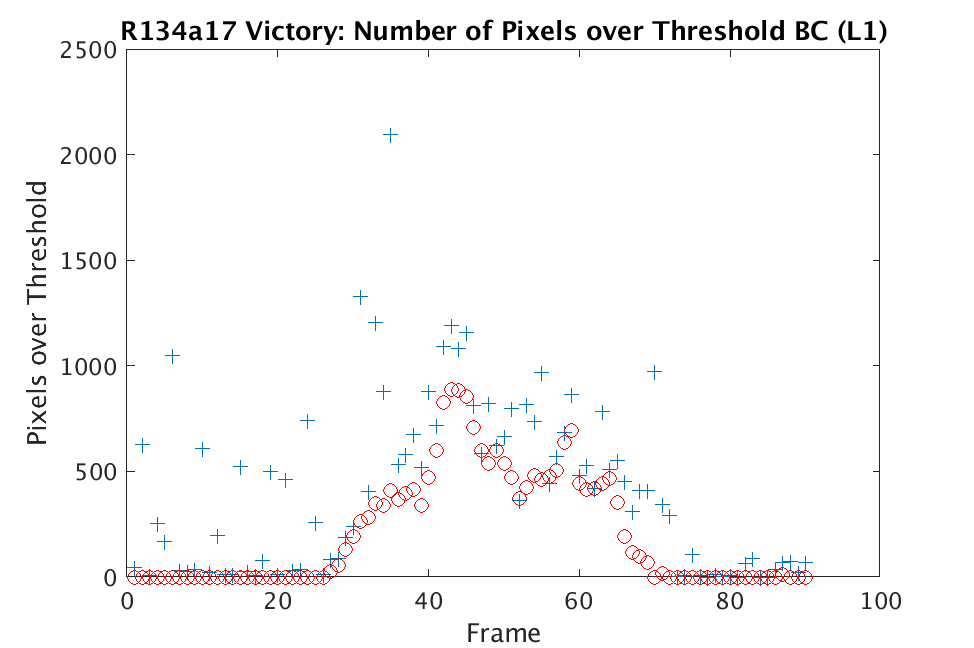}\\
\includegraphics[height=5cm]{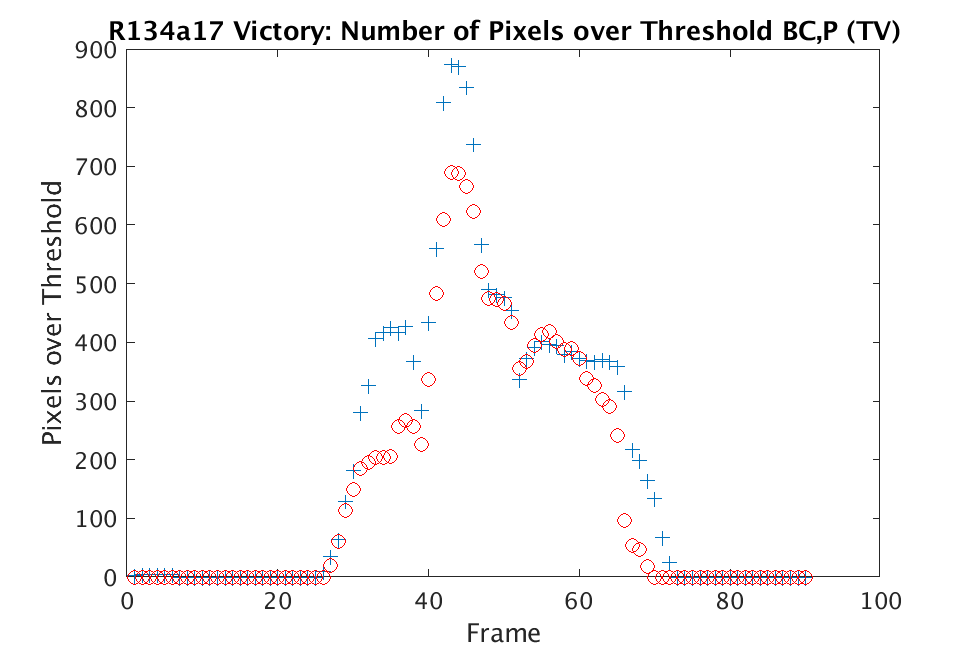}
\includegraphics[height=5cm]{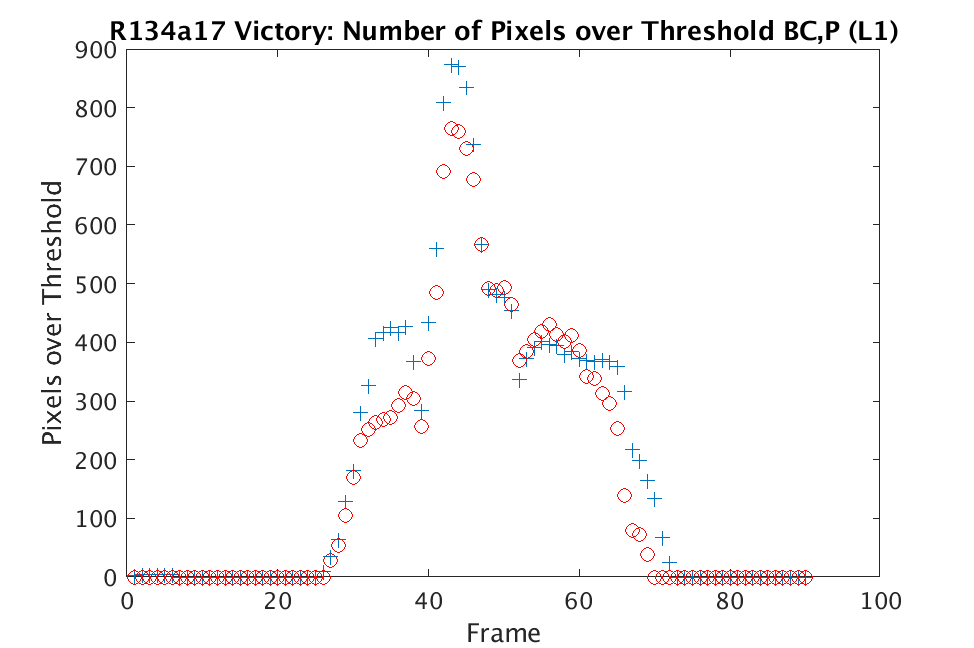}
\end{tabular}
\end{center}
\caption{\label{fig:R134a}} Comparison of the number of pixels with an ACE (bulk coherence) value that exceeds the corresponding threshold (defined as in Sec.~\ref{sec:AlgThresh}) for uncompressed data (blue crosses) and reconstructed data (red circles) as a function of time frame in the Johns Hopkins R134a 17 Victory dataset. The figures in the left column are produced from TV reconstructions, and the figures in the right column are from $\ell_1$ reconstructions. The figures are organized by row according to the number of pixels with values over the threshold for (top to bottom): ACE values, bulk coherence values, and bulk coherence with persistence values. All results are computed after sampling at $90\%$ compression. In this case, both reconstruction approaches result in chemical detection that is similar to that on uncompressed data, with a very slight improvement arising from $\ell_1$-regularization.
\end{figure}

\begin{figure}[ht]
\begin{center}
\begin{tabular}{c}
\includegraphics[height=5cm]{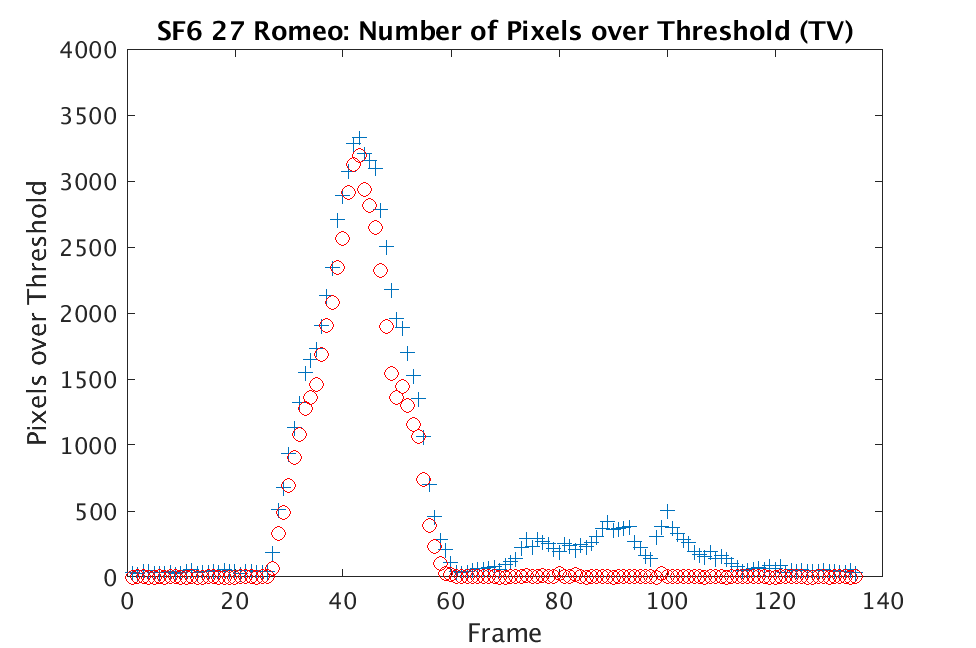}
\includegraphics[height=5cm]{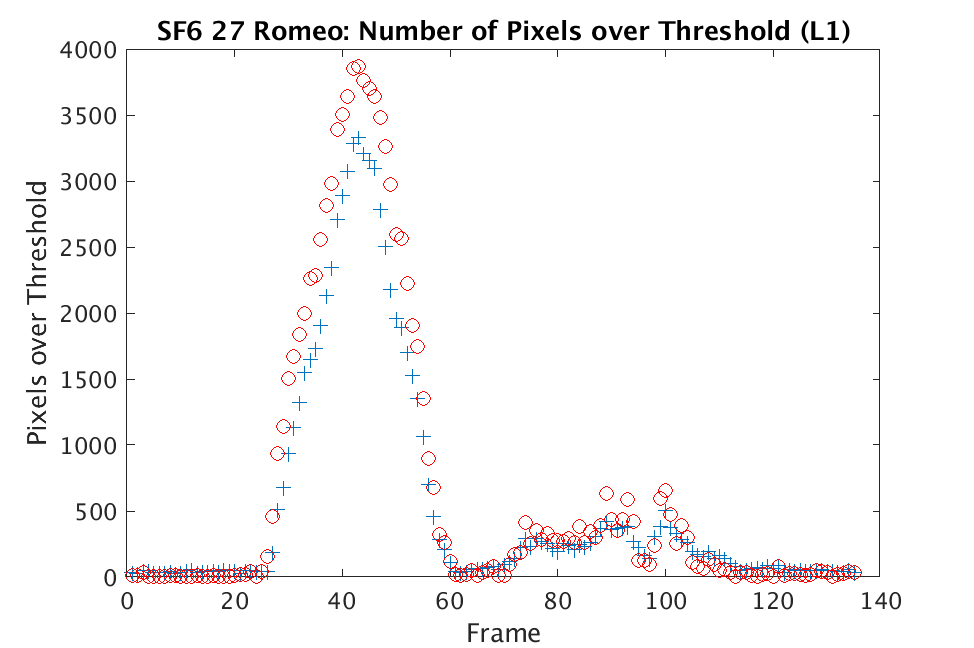}\\
\includegraphics[height=5cm]{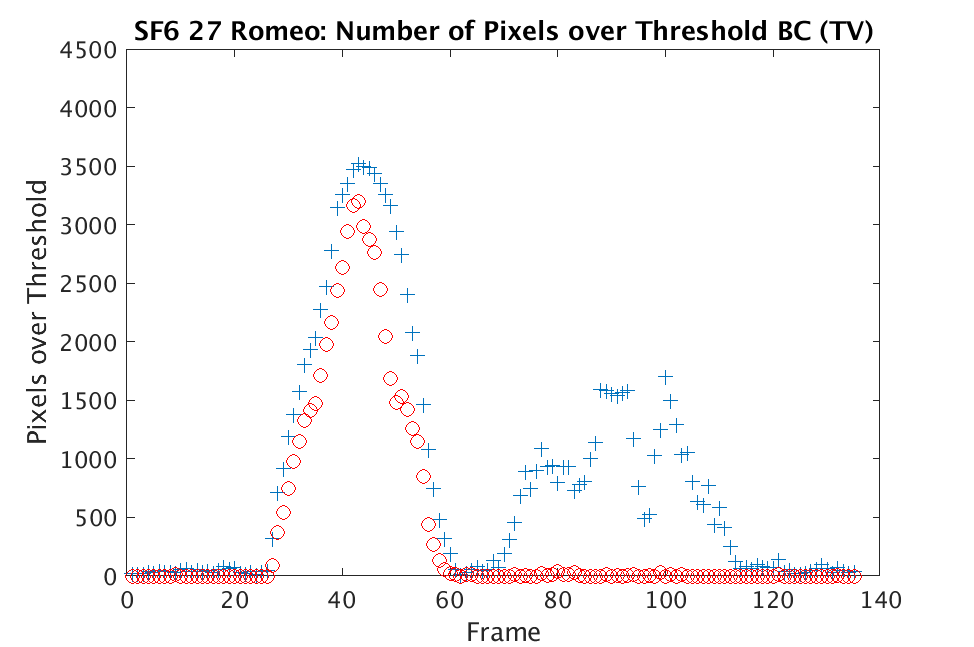}
\includegraphics[height=5cm]{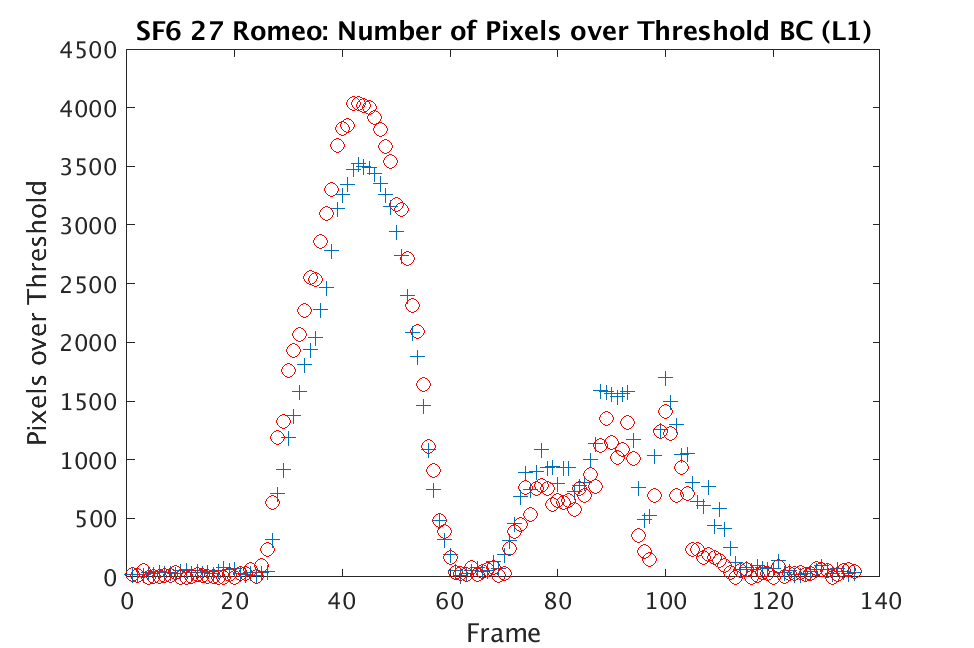}\\
\includegraphics[height=5cm]{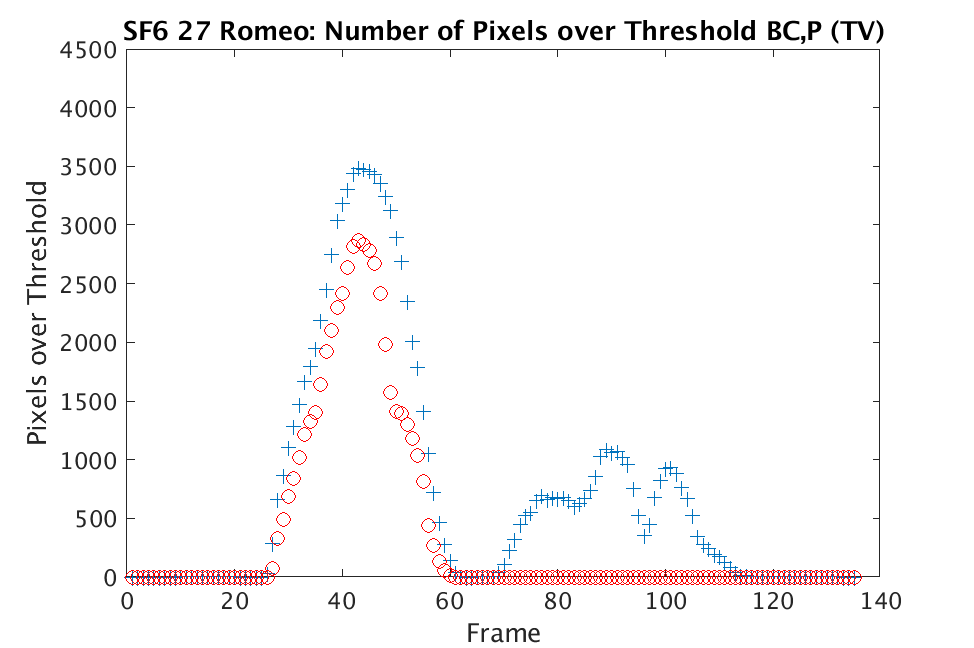}
\includegraphics[height=5cm]{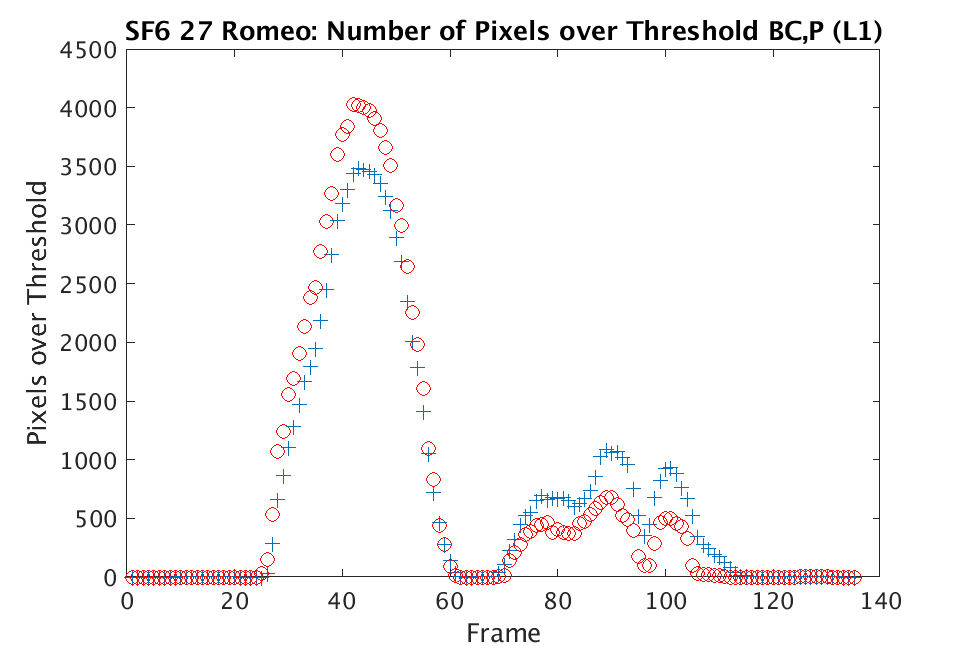}
\end{tabular}
\end{center}
\caption{\label{fig:SF6}} Comparison of the number of pixels with an ACE (bulk coherence) value that exceeds the corresponding threshold (defined as in Sec.~\ref{sec:AlgThresh}) for uncompressed data (blue crosses) and reconstructed data (red circles) as a function of time frame in the Johns Hopkins SF6 27 Romeo dataset. The figures in the left column are produced from TV reconstructions, and the figures in the right column are from $\ell_1$ reconstructions. The figures are organized by row according to the number of pixels with values over the threshold for (top to bottom): ACE values, bulk coherence values, and bulk coherence with persistence values. All results are computed after sampling at $90\%$ compression. In this case, both reconstruction approaches produce chemical detection that is similar to that on uncompressed data when the chemical is first present in the scene, with slightly stronger detection arising from $\ell_1$-regularization. However, when the chemical returns to the scene in a dissipated form, only the reconstructions produced from $\ell_1$-regularization capture the existence of the chemical.
\end{figure}

In Figs.~\ref{fig:R134a} and~\ref{fig:SF6}, we show similar results on Johns Hopkins data for R134a 17 Victory and SF6 27 Romeo. The organization of the results is the same as in Figs.~\ref{fig:GAA} and~\ref{fig:TEPA}. What is more, we observe similar behavior here: both methods appear to produce chemical detection that is consistent with that on uncompressed cubes, with slightly stronger detection occurring on the $\ell_1$-reconstructed cubes. There is one striking difference that occurs on the SF6 data displayed in Fig.~\ref{fig:SF6}. In this data, the chemical is clearly present in the field of view during time frames ~30-60, then the chemical re-enters the field of view in a weaker, dissipated form in frames ~70-110 (presumably due to a change in wind direction). The cubes reconstructed with $\ell_1$-regularization consistently detect the chemical in these later time frames, whereas those reconstructed with TV consistently fail to detect the chemical in these later frames. This dataset provides an example that speaks strongly to the recommendation of $\ell_1$-regularization over TV.

\subsection{Robustness to Threshold Variation}
\label{sec:Robustness}
We now consider the robustness of the two approaches to variation in the algorithmically determined threshold $T.$ Specifically, for each of the following datasets and corresponding chemical explosions, in addition to using $T$ to compare chemical detection, we consider a range of threshold values centered at $T$: 
\begin{equation*}
\mathcal{T}=\left[0.85T, 0.9T, 0.95T, T, 1.05T, 1.1T, 1.15T\right].
\end{equation*}
The point of these experiments is to demonstrate that it is very unlikely that our results on TV vs. $\ell_1$ are a function of the choice of threshold for detection. That is, {\emph{locally}} the ROC curve for $\ell_1$-minimization lies above the ROC curve for TV-minimization.

\begin{figure}[ht]
\begin{center}
\begin{tabular}{c}
\includegraphics[height=5cm]{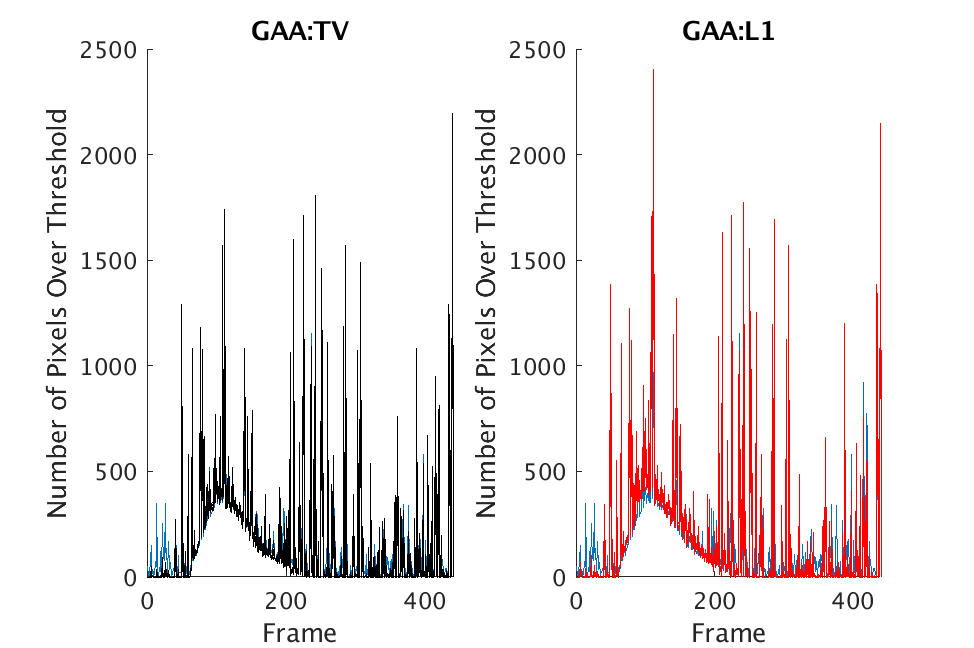}  
\includegraphics[height=5cm]{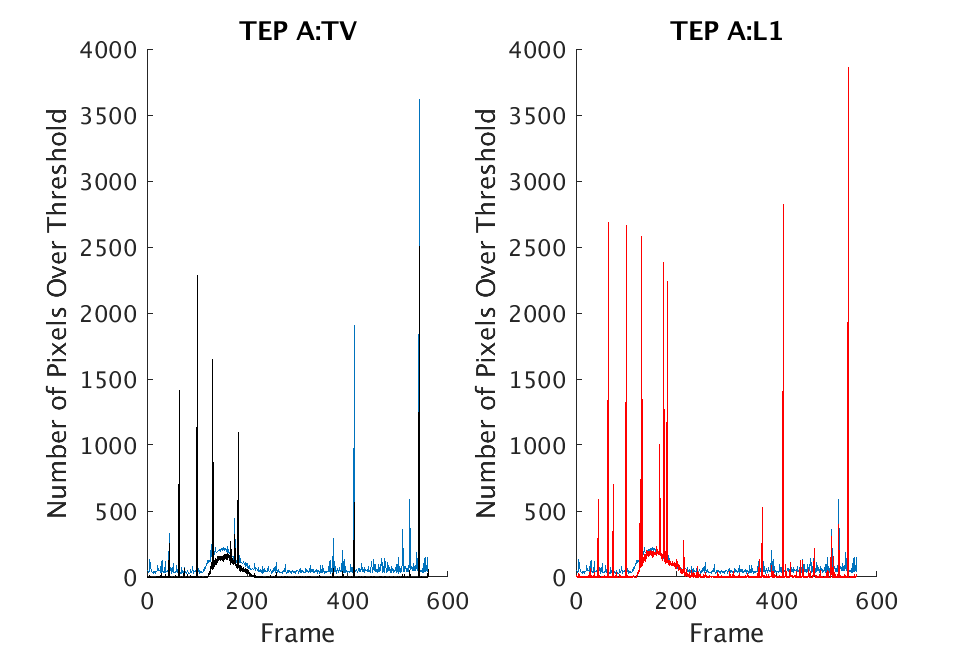} \\
\includegraphics[height=5cm]{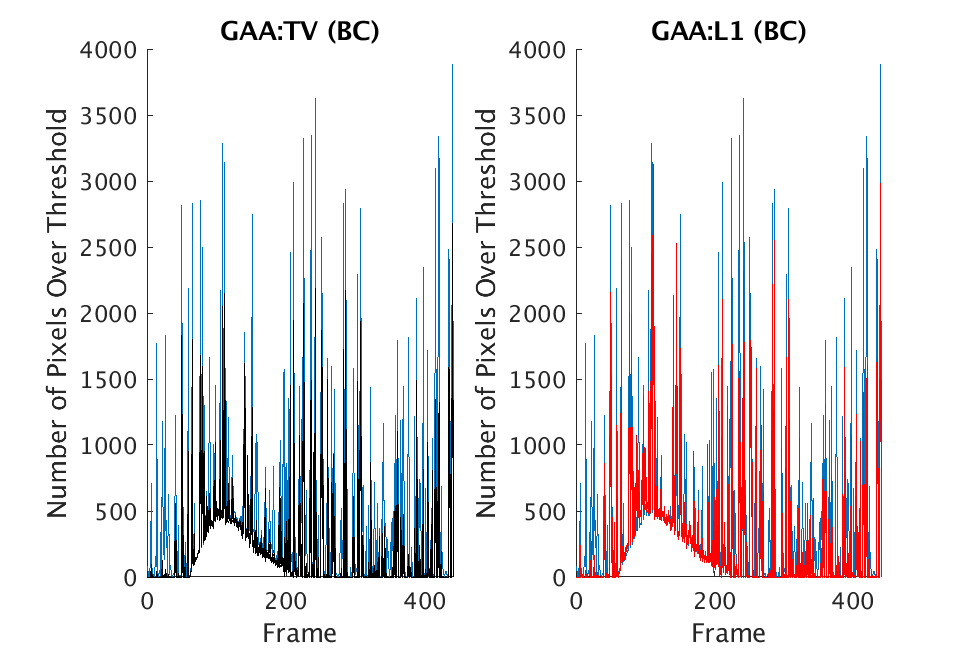}
\includegraphics[height=5cm]{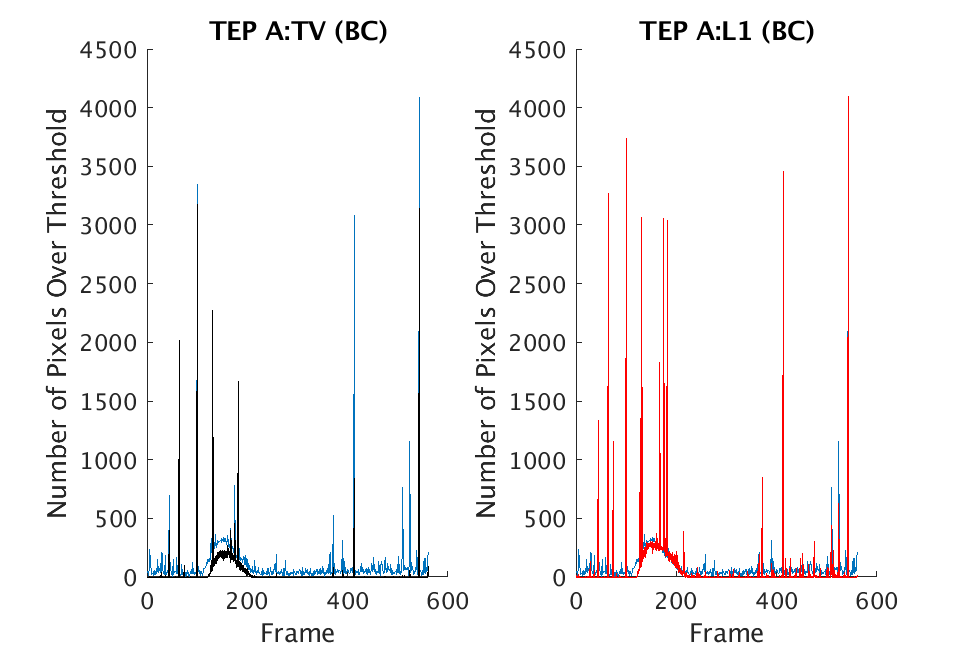}\\
\includegraphics[height=5cm]{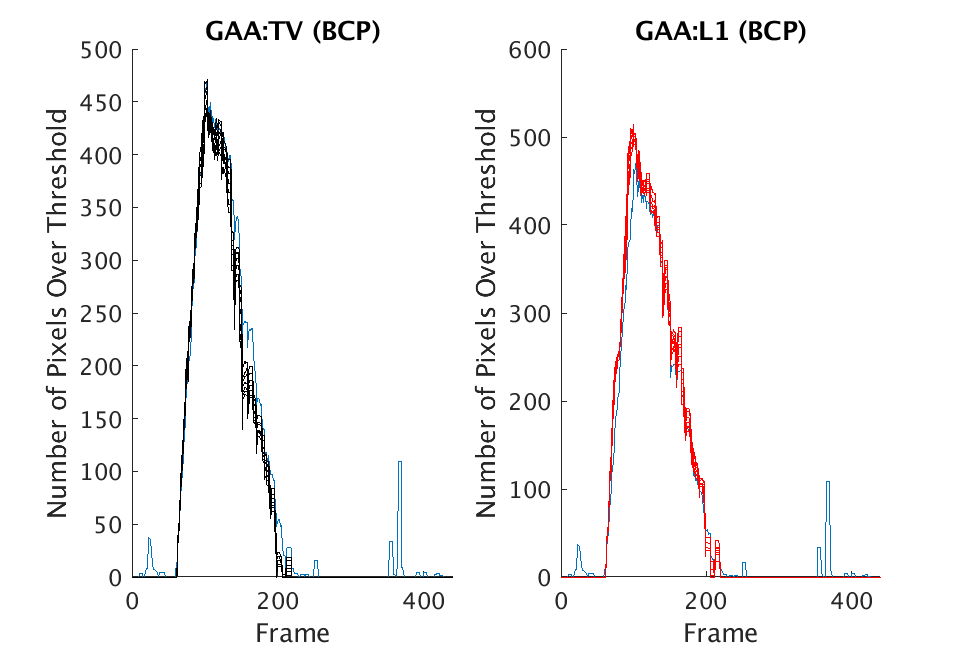}
\includegraphics[height=5cm]{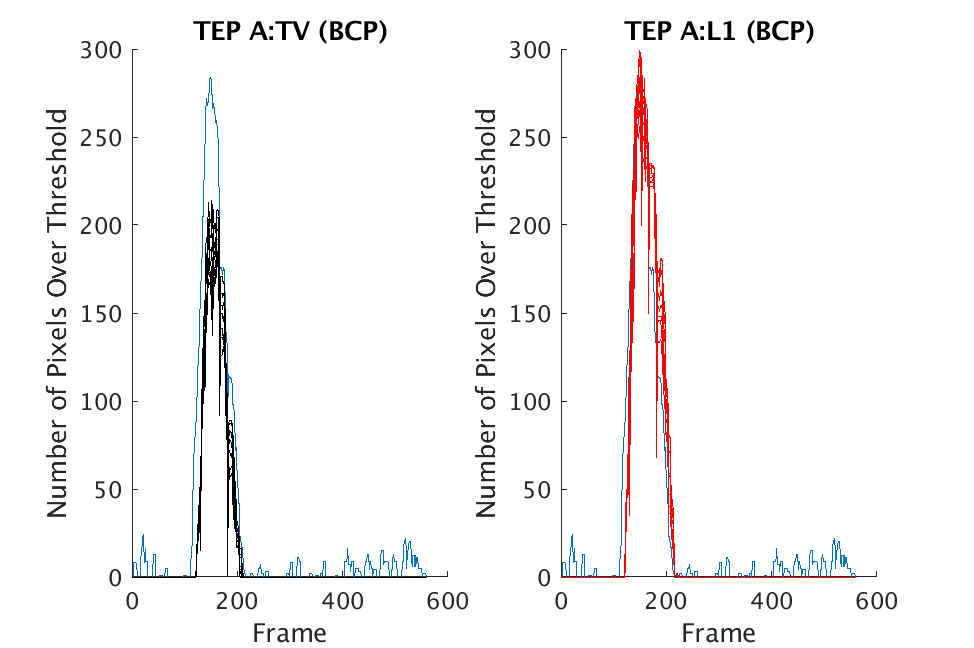}
\end{tabular}
\end{center}
\caption{\label{fig:RobustnessFP}} Comparison of robustness to threshold variation: the number of pixels with an ACE (bulk coherence) value that exceeds thresholds in $\mathcal{T}$ (as defined in Sec.~\ref{sec:Robustness}) for uncompressed data (blue curves) and reconstructed data (7 black and 7 red curves for TV and $\ell_1$, resp.) as a function of time frame in the Fabry-P\'{e}rot GAA and TEP A datasets. Results on GAA data are in columns one and two and results on TEP A data are in columns three and four. For each pair of columns, figures on the left are produced from TV reconstructions, and figures on the right  are from $\ell_1$ reconstructions. The figures are organized by row according to the number of pixels with values over the threshold for (top to bottom): ACE values, bulk coherence values, and bulk coherence with persistence values. All results are computed after sampling at $90\%$ compression. On the Fabry-P\'{e}rot data, both reconstruction approaches appear robust to variation in threshold value.
\end{figure}

We present selected results from both the Fabry-P\'{e}rot and Johns Hopkins datasets. In Fig.~\ref{fig:RobustnessFP}, we provide results on the GAA and TEP A datasets for the various thresholds for both optimization methods. All GAA data is displayed in the left column and all TEP A data is in the right column of Fig.~\ref{fig:RobustnessFP}. Blue curves are produced from chemical detection on uncompressed data and show the number of pixels over the corresponding threshold in each time frame. For each figure, there are seven curves shown in black or red that correspond to the seven thresholds considered. Note that there are artifacts and noise that lead to large spikes in the number of pixels over the threshold for both the GAA and TEP A datasets. The top row shows results for ACE, the middle row shows results for bulk coherence, and the third row shows results for bulk coherence and persistence, respectively. The cleanest comparison arises in the bulk coherence, persistence setting (note that the scales differ for the various figures). For both TV and $\ell_1,$ there are many pixels over the threshold when the chemical is present in the scene, with stronger detection occurring in the $\ell_1$ results.

\begin{figure}[ht]
\begin{center}
\begin{tabular}{c}
\includegraphics[height=5cm]{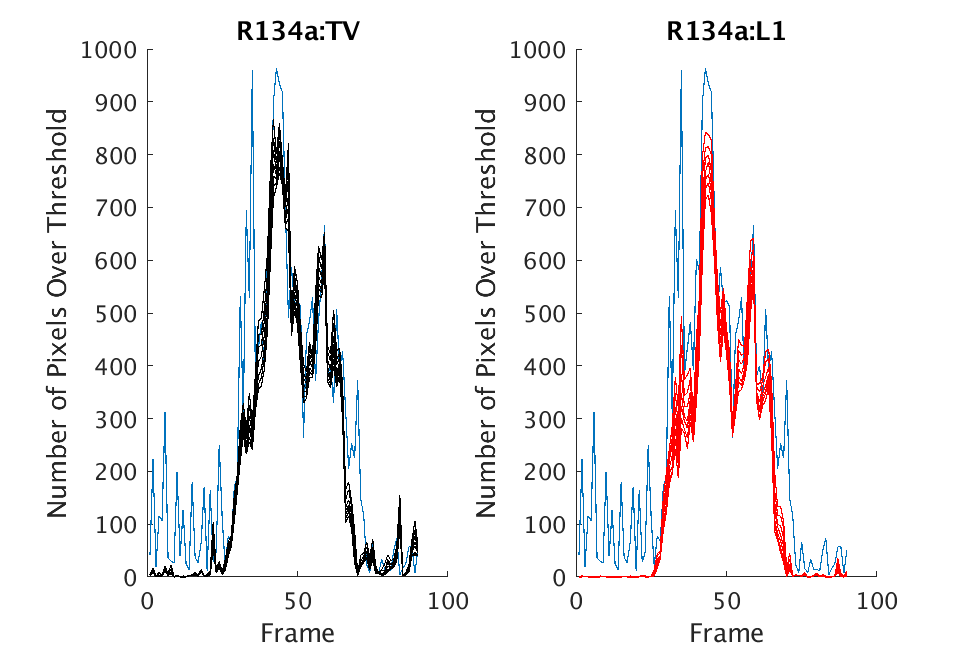}
\includegraphics[height=5cm]{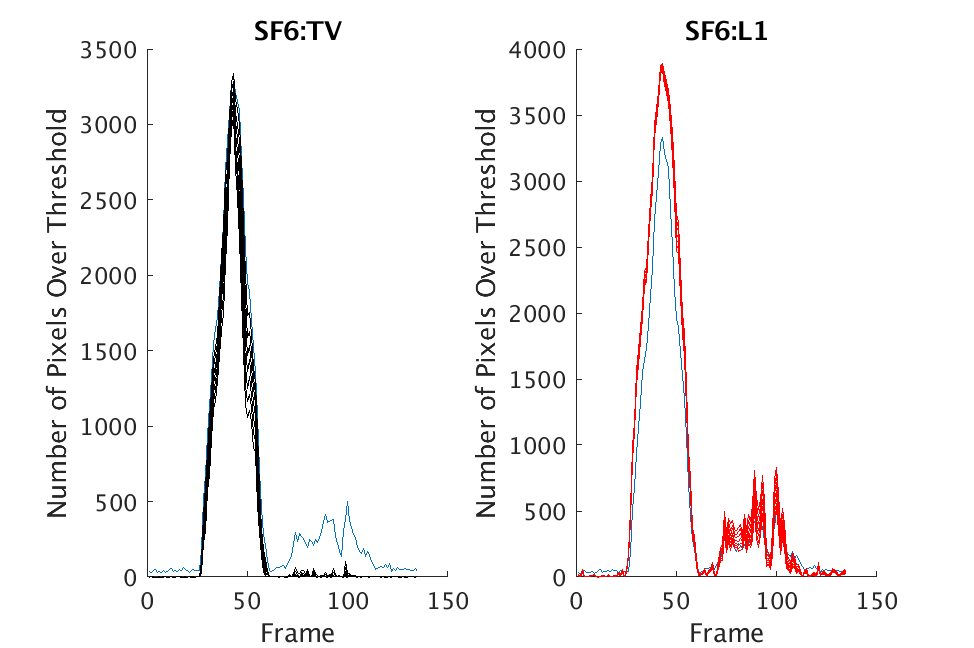}\\
\includegraphics[height=5cm]{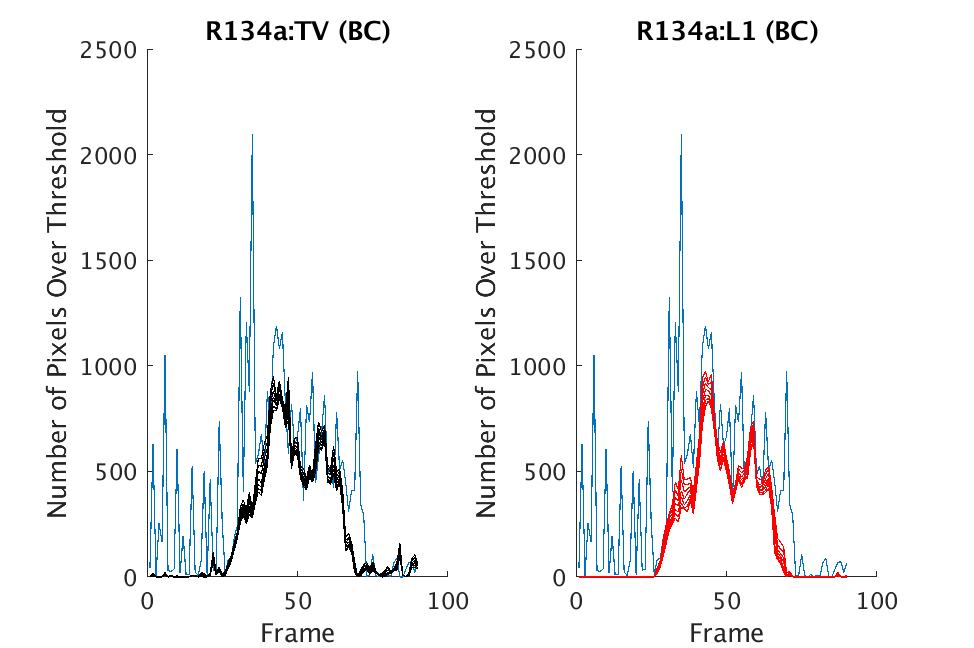}
\includegraphics[height=5cm]{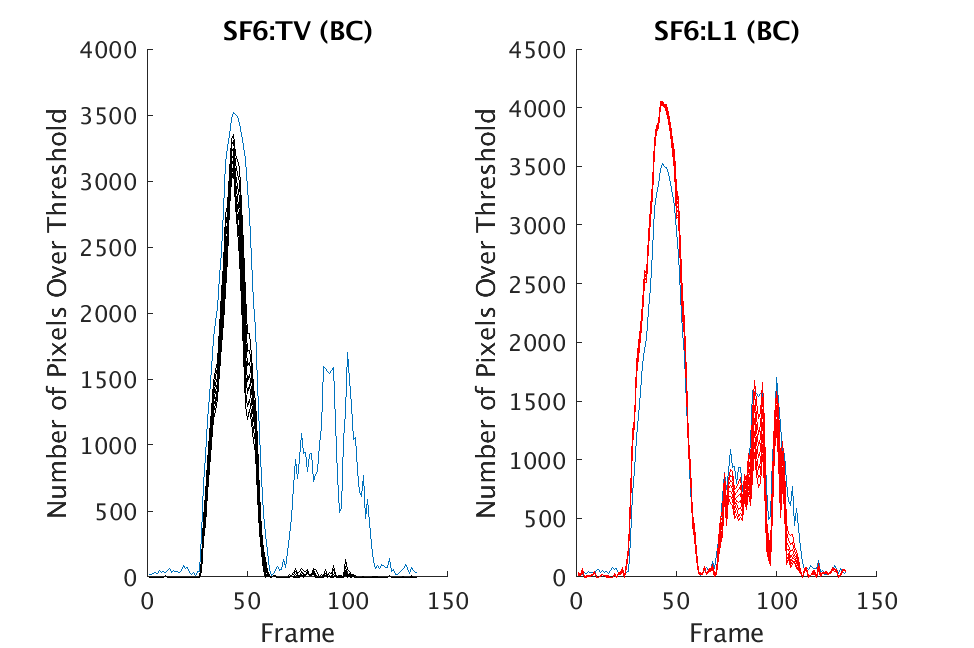}\\
\includegraphics[height=5cm]{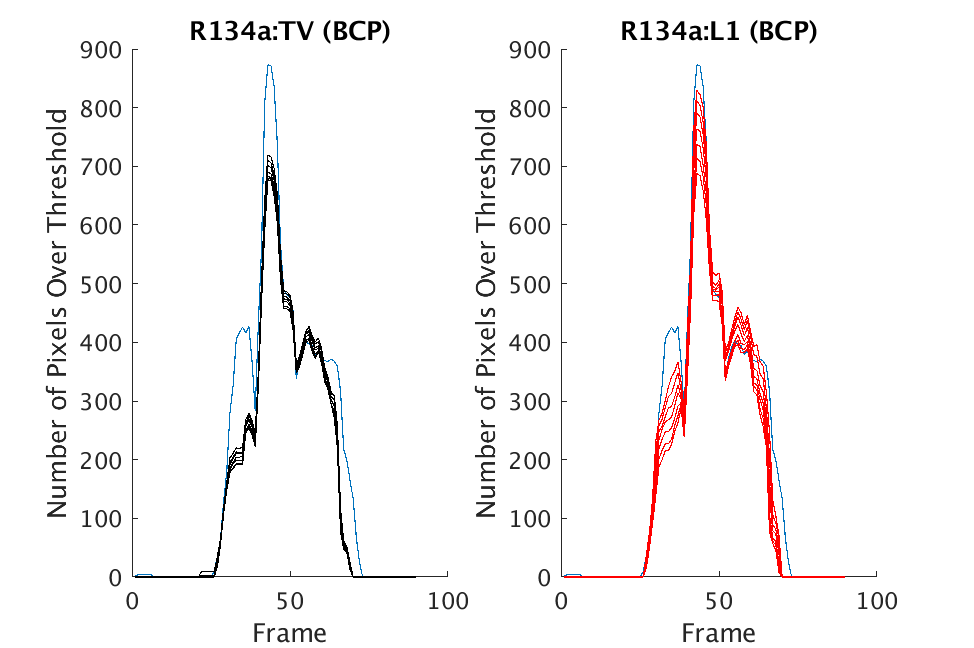}
\includegraphics[height=5cm]{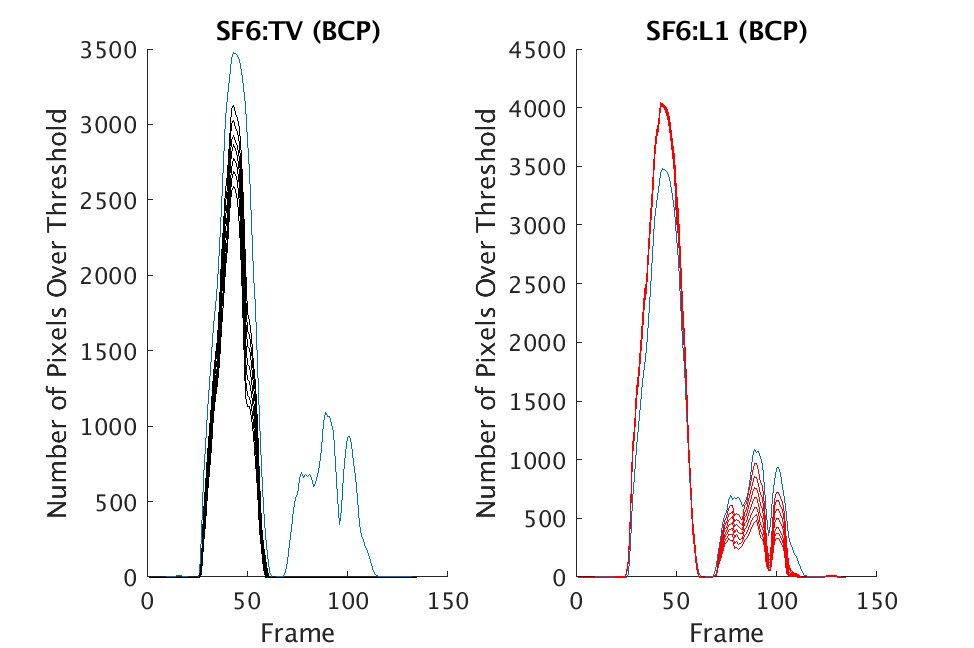}
\end{tabular}
\end{center}
\caption{\label{fig:RobustnessJH}} Comparison of robustness to threshold variation: the number of pixels with an ACE (bulk coherence) value that exceeds thresholds in $\mathcal{T}$ (as defined in Sec.~\ref{sec:Robustness}) for uncompressed data (blue curves) and reconstructed data (7 black and 7 red curves for TV and $\ell_1$, resp.) as a function of time frame in the Johns Hopkins R134a 17 Victory and SF6 27 Romeo datasets. Results on R134a data are in columns one and two and results on SF6 data are in columns three and four. For each pair of columns, figures on the left are produced from TV reconstructions, and figures on the right  are from $\ell_1$ reconstructions. The figures are organized by row according to the number of pixels with values over the threshold for (top to bottom): ACE values, bulk coherence values, and bulk coherence with persistence values. All results are computed after sampling at $90\%$ compression. On the Johns Hopkins data, we see that the level of chemical detection produced from the two reconstruction approaches remains fairly consistent. Note that even with variation of the threshold, the TV reconstruction still results in little to no detection of the return of SF6 to the scene in later time frames, whereas the $\ell_1$ reconstructions consistently detect SF6 in those same frames.
\end{figure}

Figure~\ref{fig:RobustnessJH} contains similar information, but these are results on Johns Hopkins data (R134a 17 Victory and SF6 27 Romeo). The results in Figs.~\ref{fig:RobustnessFP} and~\ref{fig:RobustnessJH} suggest that both approaches consistently produce reasonable chemical detection, with slightly stronger detection arising from $\ell_1$-regularization. The exception to this occurs in the Johns Hopkins SF6 dataset. In this case, no threshold choice in $\mathcal{T}$ leads to clear chemical detection when the chemical reenters the scene on cubes reconstructed with TV; in contrast, all choices of threshold in $\mathcal{T}$ lead to chemical detection on the same data reconstructed with $\ell_1$-regularization.

\section{CONCLUSION}
\label{sec:Conclusion}
Compressive sensing techniques are already appreciated as an important tool for hyperspectral sensing and chemical detection. We have presented evidence to make a quantitative comparison of two common approaches to reconstruction: use of an $\ell_1$-regularization term and of a TV-regularization term. We focus on a specific context, that in which the hyperspectral data has been subsampled by a specific set of linear functionals derived from rows of a Walsh-Hadamard matrix. In this setting, we have compared results on Fabry-P\'{e}rot and Johns Hopkins data. Experimentally, we have shown that, while both approaches produce good results (even at 90\% compression), optimization with $\ell_1$-regularization is generally more effective for chemical detection and more robust to threshold choices for chemical detection. 
We thus recommend $\ell_1$-regularization as a useful means of CS reconstruction of hyperspectral cubes for chemical detection. 

An avenue for further work is to demonstrate a theoretical justification for the differences in chemical detection performance on reconstructed hyperspectral cubes. Other questions that would be useful directions for future work include the following: how does the choice of sampling matrix affect the comparative performance of $\ell_1$ and TV? Are these results consistent across different levels of compression? How does the choice of sparsity promoting basis affect performance?

\acknowledgments 
The authors would like to thank Louis Scharf for insightful discussions related to this work, especially with regard to content involving ACE and MPACE. This research was partially supported by 
Department of Defense Army STTR Compressive Sensing Flash IR 3D Imager contract W911NF-16-C-0107
and Department of Energy STTR Compressive Spectral Video in the LWIR contract W911SR-17-C-0012.
\bibliography{report} 
\bibliographystyle{spiebib} 

\end{document}